# Electronic Structure of Single-Walled Carbon Nanotubes Governed by Odd-Electrons Interaction

ELENA F. SHEKA[*], LEONID A.CHERNOZATONSKII[**]

[*]*Peoples' Friendship University of the Russian Federation, 117198 Moscow, Russia.* sheka@icp.ac.ru
[**]*Institute of Biochemical Physics, Russian Academy of Sciences, 119991 Moscow, Russia.* cherno@sky.chph.ras.ru

**ABSTRACT**: For the first time, an approach is suggested for the quantitative description of the electronic structure of single-walled carbon nanotubes and the prediction of active sites for the tube controlled functionalization in view of the tube length and ends structure. The approach is based on the tube characteristic feature connected with odd electrons and is illustrated for a family involving fragments of arm-chair configured (4,4) single-walled nanotubes differing by length and end structure. Short and long tubes as well as tubes with cap end and open end, in the latter case, both hydrogen terminated and empty, are considered. Algorithms of the quantitative description of the tubes of any length are suggested. Calculations were performed in the framework of the single-determinant unrestricted Hartree-Fock approach.

**Key words:** single-walled carbon nanotubes, effectively unpaired electrons, Hartree-Fock approximation, quantitative structure-activity relations, chemical susceptibility.

## 1 Introduction

A growing number of physical, chemical, and biomedical applications of single-walled carbon nanotubes (SWNTs) require chemical modification of the latter to make them more amenable to rational and predictable manipulation. Two main strategies concern electronically weak (non-covalent) [1] and strong (covalent) [2] functionalizations have been elaborated and have gained large practical success and achievements (see comprehended reviews [3-8]). However, controllable functionalizing techniques which can provide SWNTs tailoring in a determinable manner are far from completing that still requires more investigation to elucidate the nature and locality of covalently attached moieties. Due to extreme complication of native SWNTs as well as difficulties in their separation and individual studying, the main load in looking for answers is put on shoulder of computational approaches. Starting in [4], a number of calculations have been performed [9-12].

Until now the main concept of the connection between the electronic structure and chemical reactivity of SWNTs was based on the Haddon approach [13] that attributed the reactivity of fullerenes to enormous strain engendered by their spherical geometry as reflected in pyramidalization angles of carbon atoms. Just as in the case of fullerenes, curvature-induced pyramidalization and misalignment of the π-orbitals of the carbon atoms [14-16] induces a local strain in a defect-free SWNT. This concept has made allowance for explaining the difference between the reactivity of a SWNT endcap and sidewall in favor of the former, and it has given a simple explanation of the reactivity increase while the SWNT diameter decreases. Further development of the approach turned out to be useful for concerning the reactivity of the convex and concave side walls of SWNTs towards addition reactions [9].

The Haddon approach is quite productive but basically empirical while there is a possibility to consider the problem of SWNT reactivity from another standpoint better grounded theoretically. The matter is that SWNTs belong to a particular class of species with odd electrons alongside with aromatic compounds, fullerenes, graphite, graphene, etc. The nomination "odd electrons" was introduced when describing the electronic structure of diradicals. With respect to SWNTs, the term indicates that the number of valence electrons of each carbon atom is larger by one than the number of interatomic bonds it forms. Nevertheless, the problem concerning these electrons has not been raised until it has been recently shown that the electrons behavior plays a crucial role in peculiarities of the fullerene electronic structure in general and molecular chemistry in particular [17-22].

The main issue of the odd-electron concept concerns a new view on chemical bonding. The traditional theory of chemical bonding has taken conceptual and quantitative determination within the framework of a single-determinant close shell restricted Hartree-Fock (RHF) approximation [23]. Addressing to odd-electron systems, this corresponds to a limit case of strong coupling between the electrons. The approach generalization for systems with arbitrary coupling can be reached by taking the odd electrons spin into account. Within the single-determinant approximation, the open shell unrestricted Hartree-Fock (UHF) approach allows for introducing a quantitative characteristic of the odd electron interaction which is determined by exchange integrals $J = J_0 + K$, which involve both Coulombic and exchange interaction [24, 22].

If $J$ is big ($|J| > |J_{bord}|$), UHF solution corresponds to a pure spin (PS) one and coincides with RHF solution. Both solutions are stable and coupling between the electrons converges into a conventional strong covalent bonding. If $|J| < |J_{bord}|$, both HF solutions become unstable. The UHF energy is lower than that of the RHF and, consequently, UHF solution is more stable than RHF one. However, its energy is higher than that of PS solution, so that UHF solution becomes spin-contaminated with increasing admixture of high spin states at lowering $J$. In comparison with RHF solution, the electron coupling deviates from traditional covalent bonding due to a partial and/or complete exclusion of odd electrons from the covalent bonding. Thus *effectively unpaired electrons* (EUPEs) [25] appear. The exclusion becomes more and more pronounced when the difference of the RHF and UHF energies increases. Estimates making for aromatic compounds shown that bordering value $J_{bord}$ constitutes ~-4 kcal/mol (-4.3 for penthacene) while $J$ values for $C_{60}$ and $C_{70}$ are of -1.86 and -1.64 kcal/mol [26], respectively Similar value should be expected for SWNTs as well.

Whilst the single-determinant UHF approach does not represent PS state of the odd-electron system at $|J| < |J_{bord}|$, it nevertheless provides valuable information about the state. First, this concerns the PS state energy which can be accurately determined on the basis of the obtained UHF solution following methodology suggested in [24] with a preliminary calculation of the $J$ value. Second, spin-mixed character of the UHF solution opens new possibilities for a quantitative description of chemical bonding in atomic structure within the framework of the traditional approach. If $|J| < |J_{bord}|$, an excess spin density appears in the odd-electron system [25]. Formally, the feature is a result of a non-sufficiency of the single-determinant approximation for the description of the odd-electron system. The proper approach leading to the determination of PS states requires taking into account multi-determinant configuration interaction. However, such approach forces ridding oneselves of a conventional and clear quantitative description of chemical bonding. Within the traditional consideration of the problem, the excess spin density of UHF solution brings in a natural way to EUPEs [25], the total number of which $N_D$ increases when integral $|J|$ decreases. Partitioning the $N_D$ value over atoms forms a highly characteristic $N_{DA}$ map [20-22], which coincides with that one for free valency distribution. In a certain sense, $N_{DA}$ values highlight a hidden chemical susceptibility of the odd-electron system atoms, which later on may be disclosed in different chemical reactions. The finding gives a key to play chemistry with odd-electron systems with open eyes by performing a controllable computational synthesis of different adducts just following indication of chemical susceptibility of the system atoms presented by the corresponding $N_{DA}$ map.

Using two fragments of (4,4) defect-free and (4,4) (5-7)defect SWNT as example, it was shown [27], that $N_{DA}$ values appear and follow in synchronism with the excess of C-C distance which separates two odd electrons over a limit value $R_{cov}$ of 1.395Å. The limit corresponds to the largest C-C bond length at which a complete covalent bonding of the electrons occurs. Therefore, the $N_{DA}$ map which describes chemical susceptibility of the tube atoms, on one hand, is tightly connected with the tube structure, on the other, thus highlighting the distribution of the C-C bond length excess over the tube. This opens the way for obtaining *quantitative structure-activity relations* (QSARs) which are badly needed today not only for description of SWNTs but for evaluating risks incorporated in the application of both nanotubes themselves and other nanomaterials as well.

Leaning upon the background, we have started an extended study of (n,n) and (n,0) SWNT families (integers (n,m) determine the Hamada roll-up vector which is responsible for the diameter and helicity of defect-free nanotubes [28] as well as for their belonging to conductors, semiconductors and dielectrics). In the current paper we shall concentrate ourselves on applying $N_{DA}$ maps to quantitative characterization of armchair (4,4) SWNTs which present the main contribution into the HiPco produced SWNTs distribution [29] and for which the first observation of sidewall chemical covalent functionalization was obtained [2]. Coming papers will concern (n,n) (n=5, 6, 8, 9, 10) and (n,0) (n=8, 10, 12, 13) SWNTs. Some of these results will be mentioned in what follows.

The paper presents QSARs based on $N_{DA}$ maps related to a set of (4, 4) fragments differing by length and end structure. Short and long tubes as well as tubes with endcaps and open ends, in the latter case, both hydrogen terminated and empty, are considered. The main result concerns disclosing the possibility to divide the tube $N_{DA}$ map into a small number of characteristic standard components, each of which covers small space regions related to differently configured ends, sidewall and defects while superposition of which presents the map of the whole SWNT. The finding allows for suggesting a unique algorithm of the quantitative description of the atomically local chemical susceptibility of both defect-free and (5-7) defect (4,4)SWNT of any length and any end composition that gives a clear vision for experimentalists and engineers how to proceed with chemical covalent functionalization of (4,4) SWNTs aiming at obtaining wished chemical compositions.

## 2. Calculation details

As shown [25], within the single-determinant UHF approach the distribution of EUPEs can be described by spin density function

$$D(r|r') = 2\rho(r|r') - \int \rho(r|r'')\rho(r''|r')dr'', \quad (1)$$

where $D(r|r')$ has the form

$$DS = 2PS - (PS)^2. \quad (2)$$

Here $P = P^\alpha + P^\beta$ is the density matrix and $S$ is the orbital overlap matrix (α and β mark different spin directions). If the UHF computations are realized via the *NDDO* approximation [30], a nonzero overlap of orbitals results in $S = I$, where $I$ is the identity matrix. The spin density matrix $D$ then assumes the form [20]

$$D = \left(P^\alpha - P^\beta\right)^2. \quad (3)$$

The elements of the density matrices $P_{ij}^{\alpha(\beta)}$ can be written in terms of eigenvectors of the UHF solution $C_{ik}$

$$P_{ij}^{\alpha(\beta)} = \sum_{k}^{N^{\alpha(\beta)}} C_{ik}^{\alpha(\beta)} * C_{jk}^{\alpha(\beta)}., \quad (4)$$

where $N^\alpha$ and $N^\beta$ are the numbers of electrons with spins α and β, respectively. As was shown [20], the EUPEs number (local density) on atom *A* can be calculated as

$$N_{DA} = \sum_{i \in A} \sum_{B=1}^{NAT} \sum_{j \in B} D_{ij}, \quad (5)$$

while the total EUPEs number $N_D = \sum_A N_{DA}$ is determined as

$$N_D = \sum_{i,j=1}^{NORBS} D_{ij}, \quad (6)$$

where the summation is done over all orbitals.

The total EUPEs number can be obtained alternatively from the $(S**2)^{(1)}$ value following the relation [31]

$$N_D = 2\left(S^2 - \frac{(N^\alpha - N^\beta)^2}{2}\right), \tag{7}$$

where, according to [32],

$$S^2 = \frac{(N^\alpha - N^\beta)^2}{4} + \frac{N^\alpha + N^\beta}{2} - \sum_{i,j=1}^{NORBS} P_{ij}^\alpha * P_{ij}^\beta. \tag{8}$$

Equations (6)-(8) were used in calculations performed in this work. Results given below have been obtained by using UHF version of the AM1 semiempirical method implemented in the CLUSTER-Z1 program (see the program detailed description in [33]). The method is one of the best implementations of the single-determinant HF approach. Transparent theoretical algorithms, Slater function basis, well theoretically grounded parameterization implied in calculations of electron interaction integrals and perfectly carried out for atoms of the first two rows of the Mendeleev table make the calculation results quite accurate and reliable. Together with high efficacy of computational schemes, these advantages of semiempirical calculation techniques make them quite superior when applying to nanosize systems. The CLUSTER-Z1 program provides stable UHF calculations in singlet states supplemented by the calculations of all above-mentioned quantitative characteristics related to unpaired electrons [20].

A particular attention should be given to initial guess to both $P^\alpha$ and $P^\beta$ matrices when performing UHF calculations in the singlet state. An obvious choice in setting these matrices to zero and using $H^{core}$ as an initial guess to both matrices is followed by producing identical orbitals for α and β spins [23]. To avoid the complication, the relevant matrices were obtained in due course of one-point calculations in the triplet state related to the object starting structure and then were taken as the matrices initial guess through over the calculations performed.

### 3. (4,4) CWCNT family

A full set of the equilibrium structures of the fragments related to the studied (4,4) SWNT family and obtained in due course of the structure complete optimization is presented in Figures 1 and 2. Numbered from NT1 to NT9 the studied fragments form three groups. The first group (Figure 1) involves three fragments related to short defect-free (4,4) SWNTs differing by the end structure of the same tube sidewall. One end of all fragments is capped while the other is either open but hydrogen terminated (NT1) or empty (NT2), or capped (NT3). Attached hydrogens are not only service terminators but present a real hydrogenation of the tube open end. The collection is added by a defect (4,4) SWNT (NT4) with a pair of pentagon-heptagon defects [27]. The sidewall of NT1, NT2, and NT4 consists of *12* eight-atom rows and is completed by two additional rows of 4 and 2 atoms which form the cap ([2+4+*12*x8] configuration). The sidewall of NT3 fragment consists of *11* eight-atom rows and is completed by two additional 2+4 configurations from each tube ends. The second group (Figure 2) joins fragments NT5 and NT6 which present capped-end & open-end (4,4) SWNT but differing by the termination of the open end. They differ from fragments NT1 and NT3 by elongation by four atom rows. The third group (Figure 2) combines NT7, NT8, and NT9 fragments of (4,4) SWNT with both open ends but differing by hydrogen termination. The fragment sidewall is longer by one atom row comparatively with that of the second group. Taking together, the fragments allow for considering the following structure effects on the $N_{DA}$ values and their distribution caused by i) cap end; ii) open end hydrogen terminated or otherwise; iii) a pair of pentagon-heptagon defects; iv) the tube sidewall elongation.

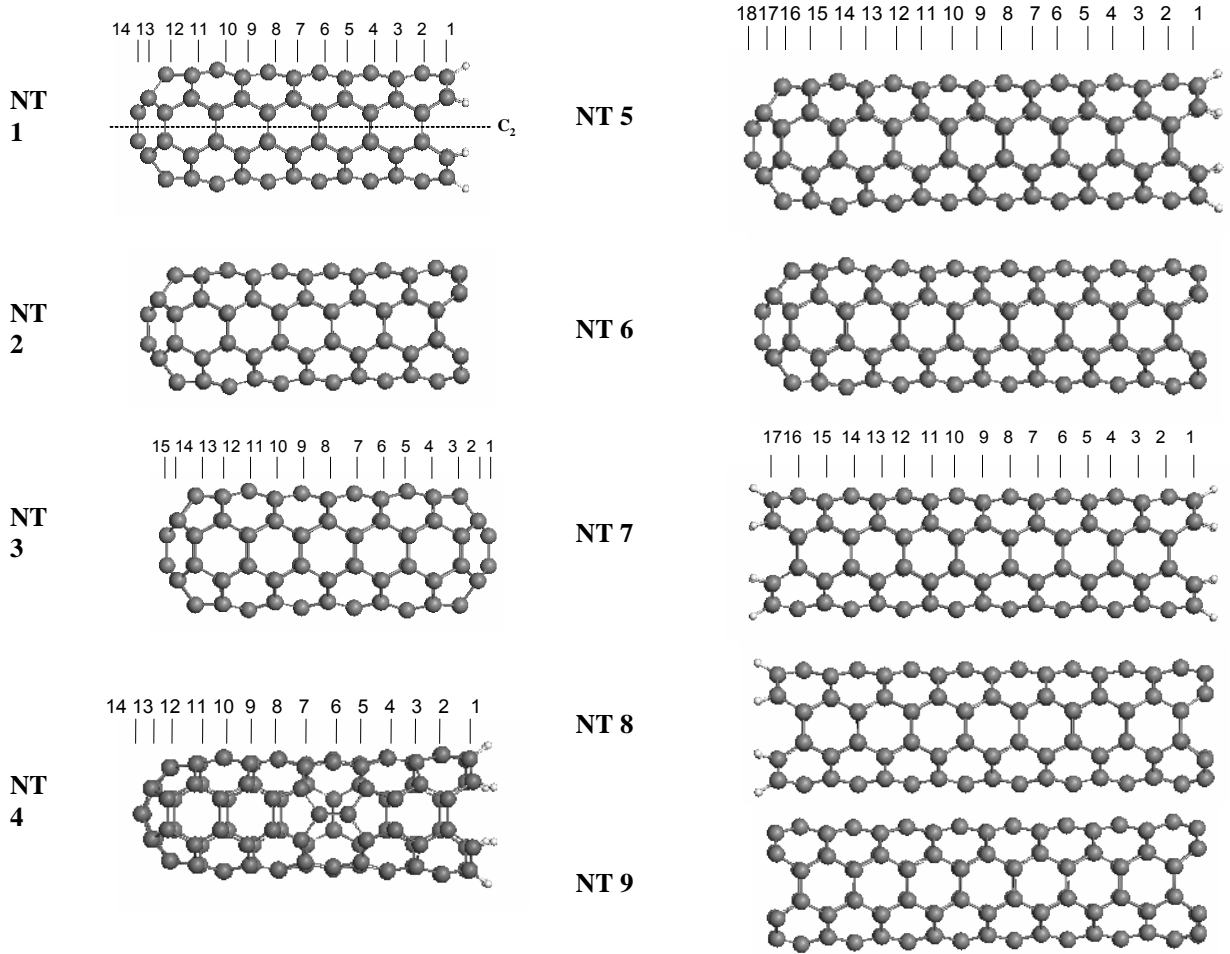

Figure 1. Equilibrium structures of (4,4) SWCNT fragments. Group 1. UHF singlet state.

Figure 2. Equilibrium structures of (4,4) SWCNT fragments. Groups 2 and 3. UHF singlet state.

### 4. Results

#### 4.1. (2+4+*12*x8) and (2+4+*11*x8+4+2) configurations, group 1.

Eight hydrogen atoms terminate the NT1 open end. The tube symmetry is $C_{2v}$. The atoms are numbered from the cap towards the open end following their symmetrical disposition with respect to both $C_2$ axis and the symmetry plane passing through the axis normally to the paper sheet. The numbering of atom rows proceeds in the opposite direction. This enumeration will be retained in what follows when presenting calculation data for other fragments.

As shown previously [27], the dominating majority of C-C bonds of the tube are longer than the limit $R_{cov}$ =1.395Å, so the large total number of EUPEs for the tube $N_D$= 32.38 does not look strange. The distribution of these electrons over the atoms forms the $N_{DA}$ map which is shown in Figure 3a. As seen from the figure, the shape of the $N_{DA}$ map supports $C_{2v}$ symmetry of the atom arrangement with accuracy not worse than 0.5-0.4%. According to the map, the tube can be divided into three regions. The first is related to the cap with adjacent atoms and covers rows 9-14. The second concerns mainly the tube sidewall and covers rows 4-8. The third refers to the open end terminated by hydrogen atoms and covers rows 1-3. The biggest non-uniformity of the $N_{DA}$ distribution is characteristic of the cap region. This region is characterized by the largest scatter of C-C bond lengths as well [27]. The latter is quite evident since the cap formation stresses the tube sidewall structure significantly. One should draw attention that the largest $N_{DA}$ values belong to atoms which are characterized by the largest C-C bond length and vice versa. As for the sidewall region, the $N_{DA}$ distribution is practically uniform with the $N_{DA}$ value scatter not bigger than 0.5% that is consistent with quite uniform distribution of C-C bond lengths as well. The $N_{DA}$ distribution in the end region is significantly affected by the hydrogenation. But in this region as well there is a direct correlation between the $N_{DA}$ values and lengths of C-C bonds. Therefore, the C-C bond length is actually a controlling factor in the distribution of the EUPEs density over atoms.

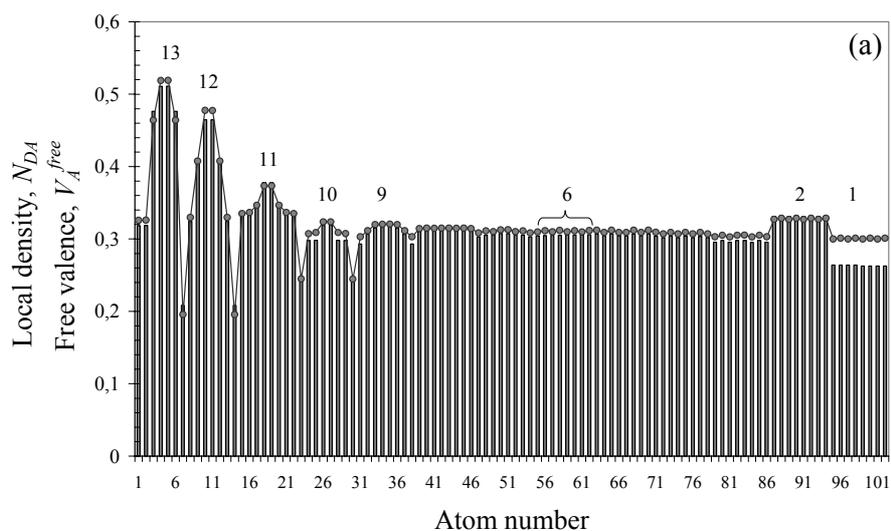
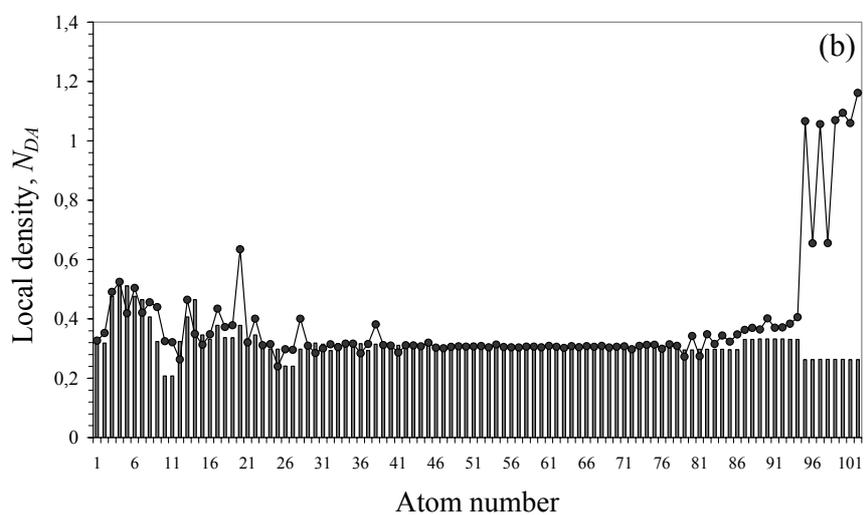
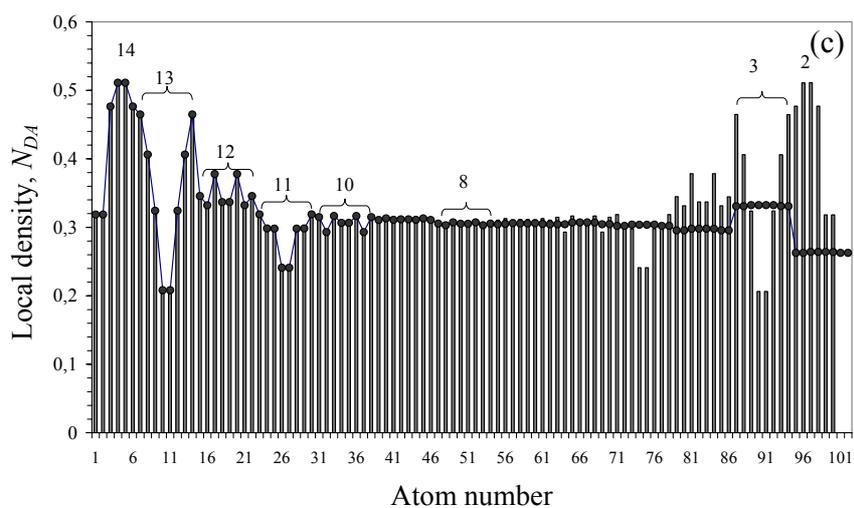

Figure 3. Local density $N_{DA}$ maps. a. NT1 fragment; bars exhibit $N_{DA}$ distribution, curve with dots presents free valency. b. NT1 (bars) and NT2 (curve with dots) fragments; c. NT3 (bars) and NT1 (curve with dots) fragments. Figures number atom rows. UHF singlet state.

The curve with dots in Figure 3a presents the calculated free valency of the tube atoms calculated in accordance with the relation

$$V_A^{free} = N_{val}^A - \sum_{B \neq A} K_{AB} \qquad (9)$$

where $N_{val}^A$ is the number of valence electrons of atom $A$ and and $\sum_{B \neq A} K_{AB}$ is the generalized bond index ($K_{AB} = |P_{AB}|^2 + |Sp_{AB}|^2$). Here, the first term is the Wiberg bond index and the second is determined by the spin density matrix (see [19] for details). The excellent agreement of the two values in the figure shows that the $N_{DA}$ map does exhibit a quantitative measure of free valency or chemical activity (susceptibility) of atoms. From this viewpoint, the tube cap is the most reactive part, while the tube sidewall is more passive with ill-pronounced selectivity along the tube.

Removing hydrogen atoms at the tube open end, we obtain the $N_{DA}$ map of NT2 fragment shown in Figure 3b. A tremendous contribution of end atoms obviously dominates on the map. It is due to the fact that the ethylene-like (*2n*) character of C-C bonds of the tube [27] is replaced by the acetylene-like (*3n*) character of the bond at the tube end where each atom has not 1 but 2 odd electrons. The transformation naturally results in increasing the total EUPEs number $N_D$ up to 39.59. The injection of additional EUPEs disturbs the $N_{DA}$ map of the hydrogen-terminated tube shown in the figure by bars quite considerably. Important to note that changes occur not only in the vicinity of the open end within rows *2* and *3* that is quite reasonable, but influences the opposite cap end (rows *14-11*). Practically no changes occur along the tube sidewall which seems to serve as a peculiar resonator for the electron conjugation. Addressing to the chemical activity of the tube, it becomes evident a dominant activity of its end atoms.

When the other end is capped, the obtained (2+4+*11*x8+4+2) configuration of fragment NT3 ($N_D$=33.35) becomes highly symmetrical ($D_{2h}$) that is reflected in its $N_{DA}$ map shown in Figure 3c. The fragment $N_{DA}$ distribution is specularly symmetrical with respect to atoms of the middle row *8*. At the same time, as seen from the figure, the addition of the second cap results in no changing in the distribution which is characteristic for the one-cap fragment NT1.

The defect introduction in the NT4 fragment was purposely made above the symmetry axis to distinctly violate $C_{2v}$ symmetry of the perfect tube. The exact tube symmetry becomes $C_s$. As seen in Figure 1, the defects are spread over rows 5-7 centering at row 6. A comparative analysis of the $N_{DA}$ maps of perfect and deformed tubes presented in Figure 4 shows that, actually, the main alterations concern atoms of these rows. The remainder part of the sidewall region is less affected. As for the cap and open end regions of the tube, the alterations also are rather weak. And nevertheless, whilst small, the changes caused by the defects touch upon all atoms of the *12*-row tube.

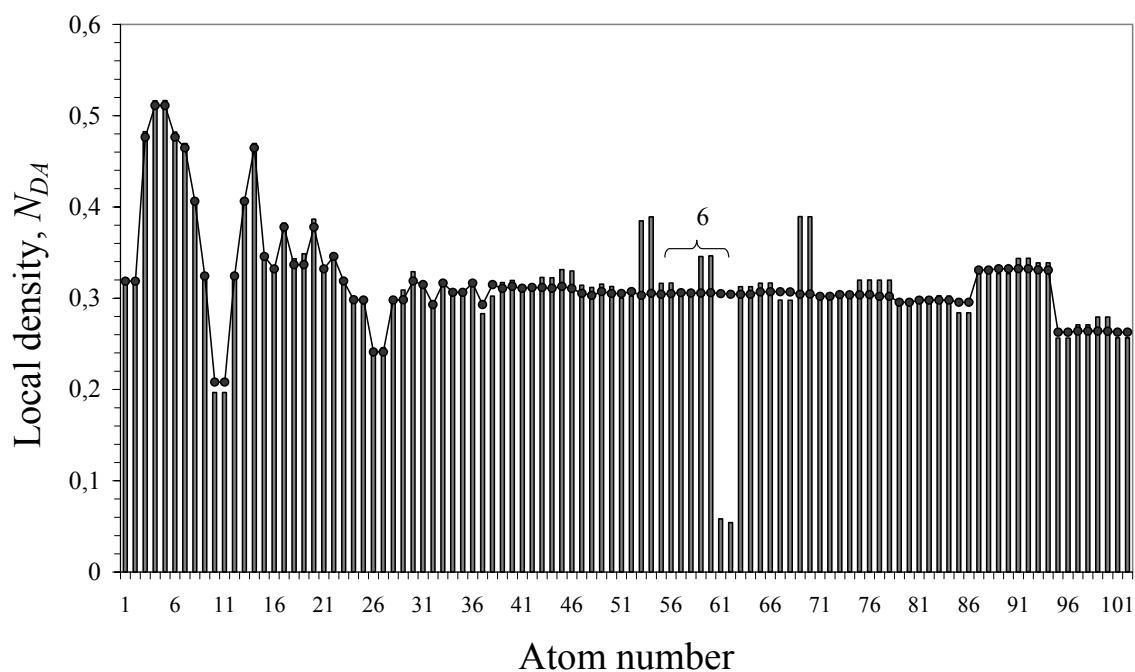

Figure 4. Local density $N_{DA}$ maps of NT1 (curve with dots) and NT4 (bars) fragments. UHF singlet state.

A significance of the open end termination can be traced through charge characteristics of the fragments as well. In the series NT1, NT2, NT3 dipole moments of fragments are 9.93 D, 0.44 D, and 0.03 D, respectively. A drastic drop of the value when terminating hydrogen atoms are removed is evidently connected with charge distribution over tube atoms. Figure 5 presents the charge distribution for the three fragments. Bars in the figure show the charge distribution along NT1 fragment with atom enumeration that coincides with that of the $N_{DA}$ map in Figure 3a. A $C_{2v}$ symmetry pattern of the distribution is clearly seen. As seen from the figure, attached hydrogens as well as adjacent carbon atoms provide the main contribution to the distribution. The influence of terminators is still sensed at another two atomic layers in the direction towards the tube cap. The remainder tube sidewall is practically uncharged, and a weak charging of atoms appears again only at the endcap. The acquired total negative charge is fully compensated by positively charged hydrogen atoms. However, significantly charged "tail" of the fragment provides a high value of its dipole moment.

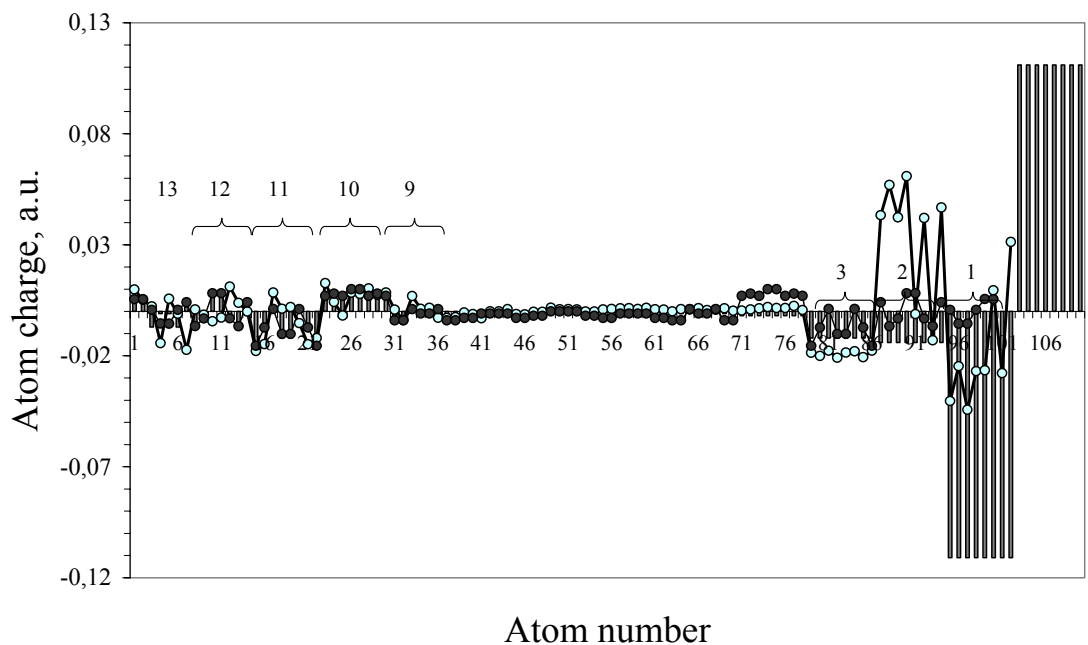

Figure 5. Charge distribution over atoms of fragments of group 1. Bars, curve with light and dark dots plot data for fragments NT1, NT2, and NT3, respectively. Figures number atom rows of fragments NT1 and NT2. UHF singlet state.

When hydrogen atoms are removed the charge map is reconstructed. First, it concerns end carbon atoms whose charge becomes much less but the charging area is still large covering rows *1-3*. Besides, the charge redistribution is seen in the cap region as well while along the tube sidewall changes are small. And again, as in the case of the $N_{DA}$ distribution, one can speak about a resonant character of the disturbance transfer along the tube sidewall. One might suggest that this very charge transfer along the tube results in so significant decreasing of the dipole moment value in spite of a still considerable charging of the tube open end.

Replacing the open end by the second cap does not violate the charge distribution in the region of the first cap and simply duplicates it in the region of the second cap just making the total tube $N_{DA}$ map specularly symmetrical with respect to the middle row of atoms. Highly symmetric and low by value, the charge distribution produces exclusively small dipole moment.

### 4.2. (2+4+16x8) configuration, group 2

$N_{DA}$ maps given for fragments NT5 and NT6 in Figure 6 show the following. Like for members of group 1 considered in the previous Section, the $N_{DA}$ map of the fragments consists of three regions related to endcap on the left, open end on the right, and an extended tube sidewall. As for both end regions, the data for elongated tube fully reproduce those for a shorter tube non depending on if the open end is hydrogen terminated (Figure 6a) or empty (Figure 6b). Basing on the results presented in the previous Section, it is possible to predict the same behavior of the map if the open end is capped. The only difference in the distributions related to shorter and longer fragments consists in expending a homogeneous part of the distributions related to the tube sidewalls. Therefore, the $N_{DA}$ map may be presented as consisting of three standard fragments MapI, MapII, and MapIII. In the case of two cap ends, MapIII should be replaced by MapI. As seen from Figure 6, MapI covers region of first (and/or last) 37-40 atoms. MapIII is spread over 24 last atoms. MapII takes the remainder space which depends on the tube length.

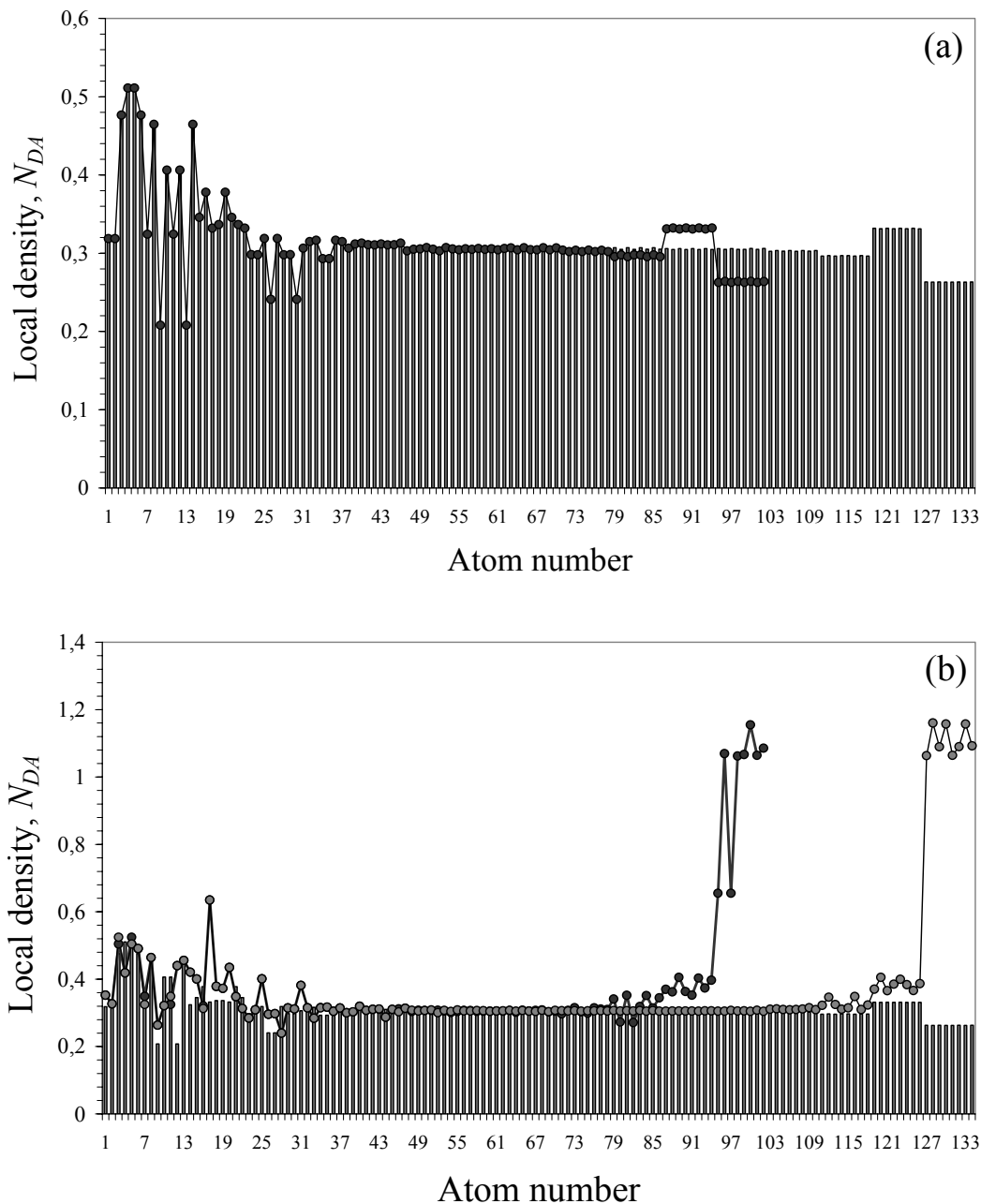

Figure 6. Local density $N_{DA}$ maps. a. NT5 (bars) and NT1 (curve with dots) fragments; b. bars, curves with light and dark dots plot data for NT5, NT6, and NT1 fragments, respectively. UHF singlet state.

### 4.3. (17x8) configuration, group 3

Group 3 covers fragments of the (*17*x8) configuration with both open ends that is consistent by length with the fragments of the (2+4+*16*x8) configuration of group 2 (see Figure 3). Similarity in the tube length allows for revealing changing caused by replacing one or two cap ends by open ends.

Figure 7a presents a comparative view on $N_{DA}$ maps of fragments NT5 and NT7 that have similar right open ends terminated by hydrogen and different left ends presented by cap in the case of NT5 and by hydrogen terminated open end in NT7. A comparison shows that if three-region pattern described above is characteristic for capped NT5 fragment, this can not be said about NT7. The $N_{DA}$ map in the case is specularly symmetrical with respect to the middle row but is quite inhomogeneous with a peculiar step-like character reaching minimum at the tube center. It should be noted nevertheless that the step character related to the last three rows on the right end is the same as for the NT5 open end.

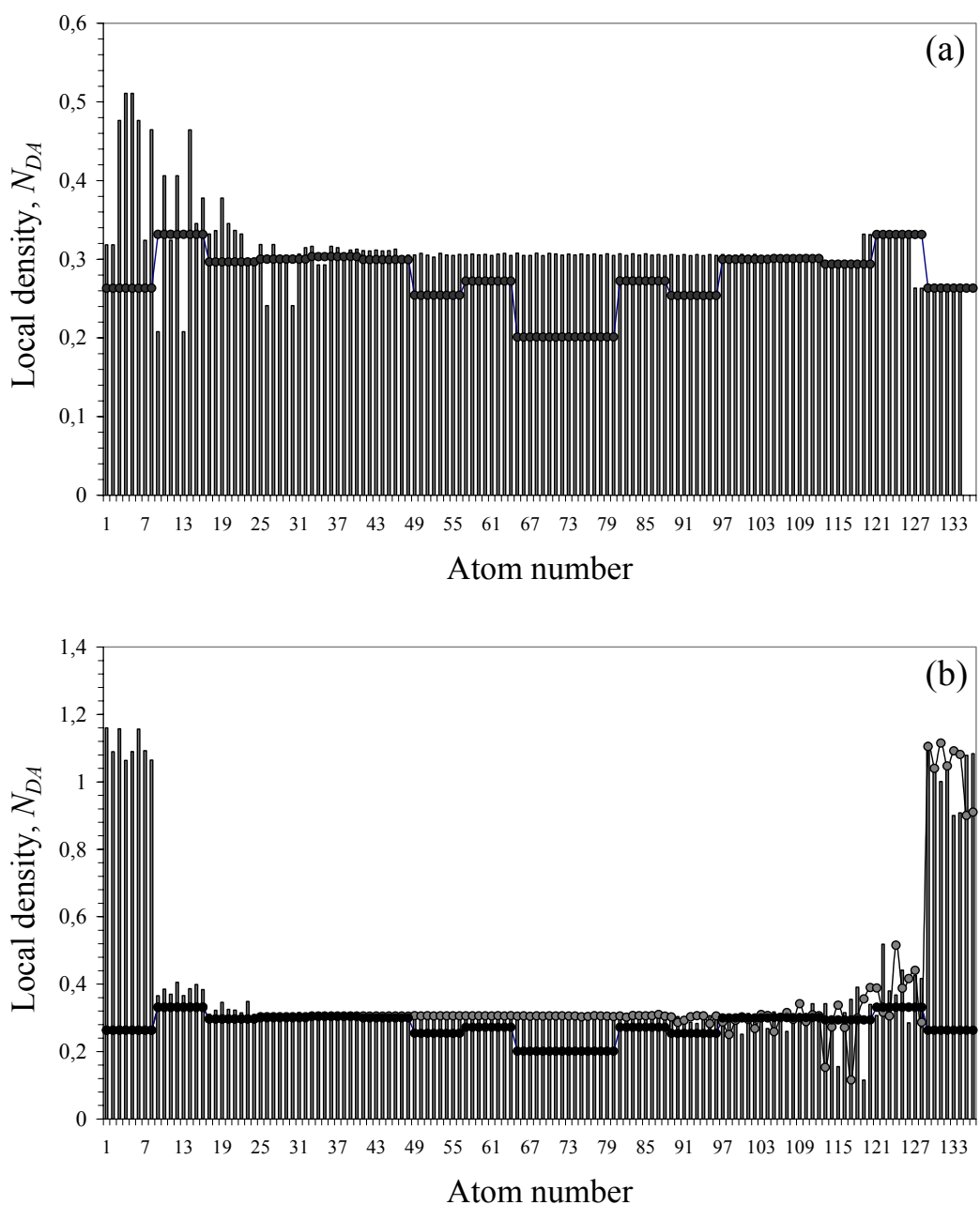

Figure 7. Local density $N_{DA}$ maps. a. NT5 (bars) and NT7 (curve with dots) fragments; b. bars, curves with light and dark dots plot data for NT9, NT8, and NT7 fragments, respectively. UHF singlet state.

Removing hydrogen atoms on the right end of the tube brings back the $N_{DA}$ map of fragment NT8 to a standard three-region form described in the previous Section (see Figure 7b) but expressed as MapIII+MapII+MapIII$^*$. MapII and MapIII have the same shape as previously while MapIII$^*$ concerns 24 last atoms after removing hydrogen terminators. When the remainder eight terminators are removed on the left end, the $N_{DA}$ map of fragment NT9 acquires a new three-region form expressed as MapIII$^{**}$+MapII+MapIII$^*$. As a whole the map becomes almost specularly symmetrical with respect to the middle row of atoms while MapIII$^{**}$ somewhat differs from MapIII$^*$. When fragment NT9 is formed from NT7 avoiding the NT8 stage, MapIII$^{**}$ is fully identical to MapIII$^*$.

## 5. Discussion and conclusive remarks

### 5.1. Comparison with experiments

A summarized view on chemical activity of the tubes, presented in details by $N_{DA}$ maps in Figures 3, 4, 6, and 7, is given in Table 1. As seen from the table, the level of the tube activity exceeds that of fullerene $C_{60}$ in all cases. Shifting the activity to either cap or empty end regions depending on the tube shape and composition may be used in looking for practical ways of the tube chemical covalent functionalization. Lack of experimental data for individual SWNTs complicates exact verification of the obtained data. However, there are some generalized observations that can be discussed in view of the performed computations.

Table 1. Synopsis of chemical activity of (4,4) SWCNT family fragments

| Fragments | Atoms with the highest chemical activity | | Atoms with the lowest chemical activity | |
|---|---|---|---|---|
| | Locality | Comparison with fullerene $C_{60}$[1] | Locality | Comparison with fullerene $C_{60}$[1] |
| **NT1** | Capped end | 1.88 | Tube sidewall | 1.13 |
| **NT2** | Open empty end | 4.28 | Tube sidewall | 1.13 |
| **NT3** | Both capped ends | 1.88 | Tube sidewall | 1.13 |
| **NT4** | Capped end and defect region | 1.88 | Tube sidewall | 1.13 |
| **NT5** | Capped end | 1.88 | Tube sidewall | 1.13 |
| **NT6** | Open empty end | 4.28 | Tube sidewall | 1.13 |
| **NT7** | Open H-term. end | 1.13 | Tube sidewall (at minimum) | 0.93 |
| **NT8** | Open empty end | 4.11 | Tube sidewall | 1.13 |
| **NT9** | Both open empty ends | 4.27 | Tube sidewall | 1.13 |

[1] Listed is the ratio $(N_{DA})_{tube}/(N_{DA})_{full}$. The maximum $(N_{DA})_{full}$ value of 0.271 related to one pair of atoms of each of six naphthalene-core fragments of $C_{60}$ molecule [21, 22] is taken as an etalon unit.

1. Studies performed with oxidatively cut SWNT caps of ~1 nm in diameter thus revealed empty open ends interacting with primarily esters and quinines [34]. Such oxidized SWNTs have been assembled on a number of surfaces, including silver [35], highly oriented pyrolitic graphite [36], and silicon [37]. High coverage densities and orientation normal to the surface have been shown; the latter being suggestive of higher degrees of functionalization at the nanotubes ends. The findings well correlate with calculated exclusive activity of the empty-open-end (4,4) SWNTs.

2. A detailed study of diameter-dependent oxidative stability [38] has shown that smaller diameter tubes are oxidized more rapidly than larger diameter tubes. The finding is consistent with our data for a series of (n,n) and (n,0) SWNTs highlighting quantitatively decreasing the tube chemical susceptibility at increasing tube diameters.

3. Frequently, intrinsic defects on SWNTs are supplemented by oxidative damage to the nanotubes framework by strong acids which leave holes functionalized with oxygenated groups such as carboxylic acid, ketone, alcohol, and ester group [39]. This and other studies (see review [8]) point to high chemical activity of the defects, that is in good agreement with our computed data.

4. Attributing endcaps of SWNTs to the chemically active regions is widely accepted [3-10]. However, as shown by calculations, this region is not always the most active since its activity can be overcome by a particular composition of the open ends.

5. There was a skepticism concerning the sidewall activity, which was considered much smaller with respect to that of fullerenes [3-10]. However producing small diameter SWNTs has removed the doubts opening the way to large scale hydrogenation [40], fluorination [41], amination [42], and a great number of other addition reactions [8, 43] involving SWNTs sidewalls.

Therefore, available chemical data well correlate with predictions to be made on the basis of the computations performed. Particularly, the available data on experimental [40] study of hydrogenated SWNTs should be noted. As shown, a significant perturbation of SWNTs structure under hydrogenation occurs. The finding seems to be understood on the basis of our data concerning the tube with both H-terminated open ends (see Figure 7). The computations clearly highlight that attaching hydrogen atoms to the most active places of the tube causes a significant disturbance of its $N_{DA}$ map and, consequently, the tube structure which is spread over large region. Obviously, the more atoms are attached, the more drastic changes occur that is observed experimentally.

### 5.2. Electronic characteristics

The studied (4,4)SWNT fragments family provides a large set of data summarized in Table 2. The analysis of the main electronic characteristics presented in the table allows for making the following conclusions.

- According to the fragments energy, endcap/H-terminated-open-end and H-terminated-open-end/H-terminated-open-end tubes are the most energetically stable.
- Removing eight hydrogen terminators from endcap/H-terminated-open-end tubes causes energetic lost of 465.7 and 486.3 kcal/mol in the case of shorter and longer tube, respectively. Those figures correspond to -58.2 and -60.8 kcal/mol (or -2.5 and -2.6 eV) which accompany the formation of a single C-H bond in these cases. The data are in good consistence with experimental [40] and calculated data [11,12] estimating the energy of formation of one C-H bond of ~2.5eV.
- Substitution of eight terminators by the second cap requires energy of 264.8 kcal/mol.

- Emptying one open end of the H-terminated-open-end/H-terminated-open-end fragment costs of 418.2 kcal/mol ( -52.3 kcal/mol per one C-H bond) while additional removing eight hydrogen terminators from the other end adds 447.3 kcal/mol (-55.9 kcal/mol) to the energetic lost. As seen, the obtained data for endcap/open-end tube and open-end/open-end tube well coincide.
- All endcap/H-terminated-open-end tubes are polarized with large dipole moment of ~10 Db while endcap/empty-open-end and endcap/endcap tubes have very low dipole moment if nothing. Practically, this circumstance is very important when applying SWNTs either to improve characteristics of non-linear optical devices based on liquid crystalline media [44, 45] or to design new hybrid materials based on SWNTs and electron donor-acceptor nanocomposites [8].
- Among two-open-end tubes, the H-terminated-open-end/empty-open-end tube is the most polarized while H-terminated-open-end/H-terminated-open-end and empty-open-end/empty-open-end tubes have small dipole moments.
- Important to note a high level of donor-acceptor characteristics of the tubes. All tubes are characterized by not too high ionization potentials and high electron affinity. The characteristic value $I_D$-$\varepsilon_A$ is in interval of 7.4-5.9 eV that meets the requirements of the formation of tightly coupled adducts involving two or more SWNTs leading to their donor-acceptor stimulated adhesion similarly to the dimerization of fullerene $C_{60}$ [46]. At the same time this explains a high acceptor ability of SWNTs in numerous donor-acceptor nanocomposites [8].

Taking together, the obtained results clearly show a high efficiency of using $N_{DA}$ maps for detailed quantitative characterization of SWNTs.

Table 2. Basic electronic characteristics of fragments of (4,4) SWCNT family [1)]

| Fragments | UHF singlet state | | | | | |
|---|---|---|---|---|---|---|
| | Symmetry | $\Delta H$, kcal/mol | $D$, D | $I$, eV | $\varepsilon$, eV | $N_D$ |
| **NT1** | $C_{2v}$ | 1182.60 | 9.928 | 9.19 | 2.19 | 32.38 |
| **NT2** | $C_2$ | 1648.34 | 0.440 | 9.71 | 3.16 | 39.59 |
| **NT3** | $D_{2h}$ | 1447.39 | 0.029 | 9.81 | 2.45 | 33.35 |
| **NT4** | $C_S$ | 1271.77 | 9.954 | 9.13 | 2.21 | 32.60 |
| **NT5** | $C_{2v}$ | 1476.6 | 10.390 | 9.16 | 2.26 | 42.17 |
| **NT6** | $C_2$ | 1962.91 | 0.428 | 9.71 | 3.14 | 50.47 |
| **NT7** | $C_{4h}$ | 1271.42 | 0.003 | 8.20 | 2.38 | 37.96 |
| **NT8** | $C_{4h}$ | 1689.58 | 10.990 | 9.19 | 2.92 | 47.36 |
| **NT9** | $C_{2h}$ | 2136.90 | 0.225 | 9.66 | 3.04 | 54.90 |

[1)] $\Delta H$, $D$, $I$, and $\varepsilon$ present the heat of formation, dipole moment, ionization potential and electron afinity, respectively.

**Appendix**

**Algorithm of the (4,4) SWNTs electronic structure**

The $N_{DA}$ maps studied in the current paper form the ground for computational covalent chemistry of SWNTs just by selecting target atoms by the largest $N_{DA}$ value similarly to computational synthesis of fullerene derivatives [20, 21]. However, one can argue that the conclusion concerns only the studied fragments limited in size. It could have been so but establishing a standard three-region pattern of the tube composition allows for suggesting a common algorithm of the construction of the $N_{DA}$ map for a (4,4) SWNT of any length (of arbitrary number $N$ of eight-atom rows) and composition of its ends. The algorithm has rather simple form Map1+Map2+Map3 and its components are given in Table 3.

Table 3. Standard components of $N_{DA}$ maps of (4,4) SWCNT fragments

| Fragments | Components[1] | | |
|---|---|---|---|
| | 1 | 2 | 3 |
| **NT1** | MapI | MapII | MapIII |
| **NT2** | MapI* | MapII | MapIII* |
| **NT3** | MapI | MapII | MapI* |
| **NT4** | MapI | MapII* | MapIII |
| **NT5** | MapI | MapII | MapIII |
| **NT6** | MapI* | MapII | MapIII* |
| **NT7** | MapIII | MapII** | MapIII |
| **NT8** | MapIII | MapII | MapIII* |
| **NT9** | MapIII** | MapII | MapIII* |

[1] The component description see in Sections 4.2 and 4.3.

Applying to different end compositions presented by studied fragments, the $N_{DA}$ map of the corresponding (4,4) SWNT can be expressed in the following form.

• Endcap/H-termin.open end tube of the (2+4+$N$x8) configuration; atom numeration proceeds from the cap end towards open end while the row numeration is opposite. Algorithm is suggested on the basis of $N_{DA}$ maps of NT1 and NT5 fragments.

Map (NT1) = MapI [2+4, $N$x8, ($N$-1)x8, ($N$-2)x8, (N-3)x8] +
   MapII [($N$-4)x8, ...,$4$x8] + MapIII [$3$x8, $2$x8, $1$x8]. (I)

Here and throughout of the paper numbers of the corresponding rows are marked by italic. The components are described in Section 4.2.

• Endcap/H-termin.open end tube of the (2+4+$N$x8) configuration. Algorithm is suggested on the basis of $N_{DA}$ maps of NT2 and NT6 fragments.

Map (NT2) = MapI* (2+4, $N$x8, ($N$-1)x8, ($N$-2)x8, (N-3)x8) +
   MapII (($N$-4)x8, ...,$5$x8) + MapIII* ($4$x8,$3$x8, $2$x8, $1$x8, 4+2). (II)

The algorithm differs from the previous one by components MapI* and MapIII* which involve changing of components MapI and MapIII after removing hydrogen terminators. The changing is standard for all studied tubes.

• Endcap /endcap tube of the (2+4+$N$x8+4+2) configuration. Algorithm is suggested on the basis of the NT3 fragment map.

Map (NT3) = MapI (2+4, $N$x8, ($N$-1)x8, ($N$-2)x8, ($N$-3)x8) +

MapII (($N$-4)x8, ...,5x8) + MapI$^*$ (*4*x8,*3*x8, *2*x8, *1*x8, 4+2).
(III)

MapI$^*$ is specularly symmetric to component MapI. As seen from all three tube configurations, compositions 2+4+*8*x8 and 2+4+*9*x8+4+2 are the least model structures needed for reproducing the main pattern of the $N_{DA}$ map of this kind (4,4) SWNT as well for providing the relevant standard map components. The data obtained for fragments considered in the current paper meet the requirement and allow for constructing the $N_{DA}$ map of a (4,4) SWNT with one or two endcaps of any length.

- Endcap/H-terminated open end tube of the (2+4+*N*x8) configuration with a pair of pentagon-heptagon defects. Algorithm is suggested on the basis of the NT4 fragment map.

Map (NT4) = MapI (2+4, *N*x8, ($N$-1)x8, ($N$-2)x8, (N-3)x8) +
MapII$^*$ (($N$-4)x8, …defect location..,5x8) +
MapIII (*4*x8,*3*x8, *2*x8, *1*x8, 4+2). (IV)

The main peculiarity of the algorithm concerns component MapII$^*$ which should be examined for each case of the defect location.

- H-terminated open end/H-terminated open end tube; atom numeration proceeds from the left end towards the right end while the row numeration is opposite. Algorithm is suggested on the basis of the NT7 fragment map.

Map (NT7) = MapIII (*N*x8, ($N$-1)x8, ($N$-2)x8) +
MapII$^{**}$ (($N$-4)x8, ...,*4*x8) + MapIII (*3*x8, *2*x8, *1*x8). (V)

A detailed description of the MapII$^{**}$ component requires a further study of tubes of different length.

- H-terminated open end/empty open end tube. Algorithm is suggested on the basis of the NT8 fragment map.

Map (NT8) = MapIII (*N*x8, ($N$-1)x8, ($N$-2)x8) +
MapII (($N$-4)x8, ...,*4*x8) + MapIII$^*$ (*3*x8, *2*x8, *1*x8). (VI)

- Empty open end/empty open end tube. Algorithm is suggested on the basis of the NT9 fragment map.

Map (NT9) = MapIII$^{**}$ (*N*x8, ($N$-1)x8, ($N$-2)x8) +
MapII (($N$-4)x8, ...,*4*x8) + MapIII$^*$ (*3*x8, *2*x8, *1*x8). (VII)

The difference in components MapIII$^{**}$ and MapIII$^*$ is described in Section 4.3. The former may be recommended as standard for both components in algorithm VII when constructing the $N_{DA}$ map of the (4,4) SWNT of any length.

As seen from the above, not less than (*7*x8) configuration should be chosen for stimulating (4,4) SWNTs with both open ends and for determination of the relevant map components. The studied fragments NT7 and NT8 fully meet the requirement and present model structure complete enough for the simulation of the tube of such kind. Still the question remains concerning fragment NT7.

Obviously, the suggested algorithms provide an approximated general description of the (4,4) SWNT tubes. However, the description accuracy is rather high, not worse than 1% that is quite enough for a reliable application of the approach in practice. The obtained data for other (n,n) and (n,0) SWNTS support the approach making it quite general.

**Acknowledgement.** The work was financially supported by Deutsche Forschungsgemeinschaft and Russian Academy of Sciences, grant no. 436 RUS 113/785.